\documentclass[american,english,aps,manuscript]{revtex4}
\usepackage[T1]{fontenc}
\usepackage[latin9]{inputenc}
\usepackage{float}
\usepackage{amsmath}
\usepackage{amssymb}
\usepackage{graphicx}
\usepackage{esint}

\makeatletter
\@ifundefined{textcolor}{}
{%
 \definecolor{BLACK}{gray}{0}
 \definecolor{WHITE}{gray}{1}
 \definecolor{RED}{rgb}{1,0,0}
 \definecolor{GREEN}{rgb}{0,1,0}
 \definecolor{BLUE}{rgb}{0,0,1}
 \definecolor{CYAN}{cmyk}{1,0,0,0}
 \definecolor{MAGENTA}{cmyk}{0,1,0,0}
 \definecolor{YELLOW}{cmyk}{0,0,1,0}
}

\makeatother

\usepackage{babel}
\begin{document}

\title{A simple model for interactions and corrections to the Gross-Pitaevskii
Equation}

\author{Hagar Veksler and Shmuel Fishman}

\address{Physics Department, Technion- Israel Institute of Technology, Haifa
3200, Israel}

\author{Wolfgang Ketterle}

\address{MIT-Harvard Center for Ultracold Atoms, Research Laboratory of Electronics,
Department of Physics, Massachusetts Institute of Technology, Cambridge,
Massachusetts 02139, USA}
\begin{abstract}
One of the assumptions leading to the Gross-Pitaevskii Equation (GPE)
is that the interaction between atom pairs can be written effectively
as a $\delta$-function so that the interaction range of the particles
is assumed to vanish. A simple model that takes into account the extension
of the inter-particle potential is introduced. The correction to the
GPE predictions for the energy of a condensate confined by a harmonic
trap in the Thomas-Fermi (TF) regime is estimated. Although it is
found to be small, we believe that in some situations it can be measured
using its dependance on the frequency of the confining trap. Due to
the simplicity of the model, it may have a wide range of applications.
\end{abstract}
\maketitle

\section{introduction}

\selectlanguage{american}%
The ground state of a\textbf{ }weakly interacting Bose Einstein Condensate
(BEC) satisfies the Gross-Pitaevskii Equation (GPE) \cite{Pita_book,PS_review}

\begin{equation}
-\frac{\hbar^{2}}{2m}\nabla^{2}\psi+U\left(r\right)\psi+Ng_{3D}\left|\psi\right|^{2}\psi=\mu\psi\label{Time independent GP0}
\end{equation}
where $U\left(r\right)$ is an external confining potential, $\mu$
is the chemical potential, $N$ is the number of atoms and $g_{3D}=4\pi\hbar^{2}a/m$
is nonlinearity strength for the $s$-wave scattering length $a$.
The wave function $\psi$ is normalized to $1$. In this work, we
use a modified one dimensional version of GPE that will be derived
in the next section. 

Despite its simplicity, the GPE describes many experiments and became
very popular in the cold atoms community. The derivation of the GPE
\cite{Pita_book,Pethick@Smith} relies on two assumptions. The first
is the mean field approximation, i.e., all atoms have the same wavefunction,
so we may write the total wavefunction $\Psi\left(x_{1},x_{2},...x_{N}\right)$
for $N$ atoms as a product of single particle wavefunctions $\psi\left(x_{i}\right)$,

\begin{equation}
\Psi\left(x_{1},x_{2},...x_{N}\right)=\prod_{i=1}^{N}\psi\left(x_{i}\right).
\end{equation}
 The second assumption is that the interaction between atoms can be
replaced by a contact interaction
\begin{equation}
V\left(\vec{r}_{1}-\vec{r}_{2}\right)=g_{3D}\delta\left(\vec{r}_{1}-\vec{r}_{2}\right),\label{eq:delta interaction}
\end{equation}
with the $\delta$ function appropriately introduced via the pseudo-potential
theory \cite{Huang_book}. In this work, the validity of this approximation
and possible situations where the approximation (\ref{eq:delta interaction})
is not justified are studied. For this purpose we remember that in
the derivation of the GPE (\ref{Time independent GP0}) the origin
of the terms nonlinear in $\psi$ is the Hartree term \cite{Pita_book,PS_review}
\begin{equation}
E_{H}=\int d\vec{r}_{2}d\vec{r}_{1}\left|\psi\left(\vec{r}_{1}\right)\right|^{2}V\left(\vec{r}_{1}-\vec{r}_{2}\right)\left|\psi\left(\vec{r}_{2}\right)\right|^{2}.\label{eq:*}
\end{equation}
If one can assume that the variation of the wave function is small
over the regime where the potential $V\left(\vec{r}_{1}-\vec{r}_{2}\right)$
is substantial, we can approximate $\left|\psi\left(\vec{r}_{2}\right)\right|^{2}$
by $\left|\psi\left(\vec{r}_{1}\right)\right|^{2}$. In this case,
the effective potential (\ref{eq:delta interaction}) can be used
(this should be done with care \cite{Pita_book,PS_review,Huang_book}
but in one dimension, it is trivial). In general, the term (\ref{eq:*})
makes the calculations more involved. In the present work we are interested
in the exploration of the qualitative difference between the ground
state energy in contact potentials where the particles can be considered
point like and realistic potentials where the range of the interaction
potential is not negligible. Neutral atoms interact via the van-der-Waals
interaction, and the extension of the potential is often comparable
to the van-der-Waals radius \cite{van_der_walls}, and related to
effective range in scattering theory. It is usually much larger than
the Bohr radius or the ``size'' of the atom. To study this effect
we introduce (in Sec II) a model potential and show how it can be
written as a one dimensional potential consisting of three $\delta$-functions
separated by a characteristic length. The middle one models the repulsion
while the outer ones model the attraction. For the sake of simplicity
we study a one dimensional model, namely we study the corrections
to (\ref{Time independent GP0})\textbf{ }in an elongated trap. We
believe that similar effects will be found also in higher dimensions
of the trap. We are interested in weakly interacting BECs at zero
temperature with a time independent harmonic trapping potential. The
ground state density (including corrections related to non-vanishing
range of interaction) is presented in Sec. III and corrections to
the energy are calculated in Sec. IV and given by Eq. (\ref{eq:tot_corE})
which is the main result of this work. The magnitude of the correction
is estimated and the results are discussed in section V.

\section{\label{sec:modified-gp-for}modified GPE for $\delta$-functions
interaction}

We would like to write a toy model for three dimensional interactions
in one dimensional trap. Let us replace the usual GPE interaction
term (\ref{eq:delta interaction}) by 
\begin{equation}
V\left(r\right)=\left\{ \begin{array}{cll}
\frac{3\left(g_{3D}+\lambda\right)}{4\pi r_{in}^{3}}\qquad & for\qquad & r<r_{in}\\
-\frac{\lambda}{4\pi r_{out}^{2}\varepsilon_{out}}\qquad & for\qquad & r_{out}<r<r_{out}+\varepsilon_{out}\\
0 &  & \mathrm{otherwise}
\end{array}\right.,\label{eq:delta shell}
\end{equation}
where $r\equiv\left|\vec{r}_{1}-\vec{r}_{2}\right|$, the coupling
constant between particles is\textbf{ $g_{3D}>0$}, and $\lambda>0$
is the strength of non-contact attraction interactions. $r_{out}$
is a length scale which determines the interaction range, while $r_{in}$
is a much smaller length scale ($r_{in}\ll r_{out}$). We take the
limit $r_{in},\varepsilon_{out}\rightarrow0$ while $r_{out}$ is
fixed. This model does not require the use of pseudo-potential and
is similar in spirit to the introduction of a pseudo-potential. The
potential (\ref{eq:delta shell}) is composed of repulsive and attractive
terms, like the van-der-Waals potential, and therefore captures the
physics of van-der-Waals interaction without giving up the mathematical
and numerical simplicity. The simplicity results of the fact that
in the limit $r_{in},\varepsilon_{out}\rightarrow0$, (\ref{eq:delta shell})
is effectively a sum of $\delta$-functions.

According to \cite{Pethick_paper}, it is possible to formulate a
modified GPE which takes the range of the pair interaction into account,
as follows. For $\lambda=0$ one finds the standard GPE (\ref{Time independent GP0}).
The contribution due to the non-vanishing size of the particles $r_{out}$
was found by Collin, Massignan and Pethick \cite{Pethick_paper} (for
earlier work see \cite{Fu}). In our case it takes the form
\begin{equation}
\begin{array}{ccl}
\Delta E_{int} & = & -Ng_{3D}g_{2}k^{2}\left|\psi\left(k\right)\right|^{2}\\
 & =N & \left(\frac{2\lambda}{3}r_{out}^{2}\right)k^{2}\left|\psi\left(k\right)\right|^{2}
\end{array},\label{eq:de_ii}
\end{equation}
where $\hbar k$ is the relative momentum of the colliding particles.
Here $g_{2}=a\left(\frac{1}{3}a-\frac{1}{2}r_{e}\right)$ where $r_{e}$
is the effective range of the interaction and the result of the calculation
in App. A (Eq.(\ref{eq:g23D})) was used. In coordinate space, the
resulting equation is \cite{Pethick_paper} 
\begin{eqnarray}
\mu\psi & = & -\frac{\hbar^{2}}{2m}\nabla^{2}\psi+U\left(r\right)\psi+Ng_{3D}\left|\psi\right|^{2}\psi+Ng_{3D}\cdot g_{2}\nabla^{2}\left|\psi\right|^{2}\cdot\psi.\label{Time dependent 2deltas GP}
\end{eqnarray}
The leading correction to the GPE does not depend on the details of
the interparticle potential, therefore we can study the effect of
the corrections in terms of our simplified potential (\ref{eq:delta shell}). 

Since the trap is one dimensional, we wish to use a one dimensional
wave function $\psi\left(x\right)$ rather than $\psi\left(\vec{r}\right)$$=\psi_{x}\left(x\right)\psi_{y}\left(y\right)\psi_{z}\left(z\right)$.
For this purpose, we integrate (\ref{Time dependent 2deltas GP})
over the transverse directions $y$ and $z$, resulting in 
\begin{eqnarray}
\mu\psi_{x}\left(x\right) & = & E_{\perp}\psi_{x}\left(x\right)-\frac{\hbar^{2}}{2m}\frac{d^{2}\psi_{x}\left(x\right)}{dx^{2}}+U\left(x\right)\psi_{x}\left(x\right)\label{Time dependent 2deltas GP-1-1-1-1}\\
 &  & +g\left|\psi_{x}\left(x\right)\right|^{2}\psi_{x}\left(x\right)+g'\cdot g_{2}\frac{d^{2}\left|\psi_{x}\left(x\right)\right|^{2}}{dx^{2}}\cdot\psi_{x}\left(x\right).\nonumber 
\end{eqnarray}
where
\begin{equation}
g=Ng_{3D}\cdot\frac{m\omega_{\perp}}{2\pi\hbar}\left[1-2g_{2}\frac{m\omega_{\perp}}{\hbar}\right]\label{eq:g1d_may}
\end{equation}
and
\begin{equation}
g'=Ng_{3D}\cdot\frac{m\omega_{\perp}}{2\pi\hbar}.\label{eq:g'}
\end{equation}
Here $\omega_{\perp}$ is the (high) frequency of the confining trap
in the directions perpendicular to the BEC line. In the present work
we consider the regime where the confining frequency $\omega_{\perp}$
is sufficiently high so that the energies are lower than the first
excited state of the transverse motion, but is sufficiently low so
that the width of the ground state in the transverse direction, $a_{\perp}$,
is much larger than the three dimensional scattering length $a$.
Different physics is expected in the opposite regime where the requirement
$a\ll a_{\perp}$ is not satisfied (see \cite{Olshanii}). Since $g_{2}$
is typically small, we are allowed to neglect terms of the second
order in $g_{2}$ and replace $g'$ by $g$ in (\ref{Time dependent 2deltas GP-1-1-1-1}).
We would like to write a one dimensional GPE with the inter-particle
potential 
\begin{equation}
V\left(x\right)=2g\delta\left(x\right)-\frac{1}{2}g\left[\delta\left(x+l\right)+\delta\left(x-l\right)\right]\label{eq:3_potential}
\end{equation}
\foreignlanguage{english}{where $l$ is the effective extension of
the inter-particle potential to be related to the parameters $r_{out}$,
$g_{3D}$ and $\lambda$ of potential (\ref{eq:delta shell}), see
App. A . The one dimensional nonlinearity constant $g$ is related
to $g_{3D}$ by (\ref{eq:g1d_may}) and (\ref{eq:g'}). Since the
density does not change much on the length scale $l$}, the GPE (\ref{Time independent GP0})
modified by the replacement $g\delta\left(x\right)\rightarrow V\left(r\right)$
of (\ref{eq:delta shell}) and eventually by $V\left(x\right)$ of
(\ref{eq:3_potential}) can be written as
\begin{equation}
\mu\psi\left(x\right)=-\frac{\hbar^{2}}{2m}\frac{d^{2}}{dx^{2}}\psi\left(x\right)+U\left(x\right)\psi\left(x\right)+g\left|\psi\left(x\right)\right|^{2}\psi\left(x\right)-\frac{1}{2}gl^{2}\frac{d^{2}\left|\psi\left(x\right)\right|^{2}}{dx^{2}}\psi\left(x\right).\label{eq:2Deltas GP}
\end{equation}
Taking the limit $l\rightarrow0$, we recover the standard one dimensional
GPE. To establish the relation with three dimensional energy correction,
we compare between Eqs. (\ref{Time dependent 2deltas GP-1-1-1-1})
and (\ref{eq:2Deltas GP}) and replace $g'$ by $g$ resulting in
\foreignlanguage{english}{
\begin{equation}
\frac{1}{2}l^{2}=-g_{2}=-a\left(\frac{a}{3}-\frac{r_{e}}{2}\right).\label{eq:lar-1}
\end{equation}
}Using more realistic interaction functions (for example, a continuous
potential) generalize and replace the coefficient $l^{2}$ in (\ref{eq:2Deltas GP})
by a model dependent constant $g_{2}$. Hereafter, we consider only
the simple model (\ref{eq:3_potential}). Nevertheless, our results
are valid for any short range interaction. In other words, we demonstrate
the dependence of the corrections on the range of the inter-particle
potential.

Here, we add an extra term of the order of $\mathrm{const}\cdot a^{3}k^{2}$
to the standard GPE (see for example (\ref{eq:de_ii}) where $r_{out}$
is of order $a$ and $\lambda$ is of the same magnitude as $g$ which
is proportional to $a$). Note that taking into account contributions
from components of higher angular momentum in the partial waves expansion
will also add extra terms to the GPE. The magnitude of these terms
is of the order of 
\begin{equation}
E_{\mathcal{L}}\sim\mathrm{const}\cdot a^{2\mathcal{L}+1}k^{2\mathcal{L}}
\end{equation}
as derived in App. B of \cite{Huang_book}, where $\mathcal{L}$ is
the quantum number of angular momentum. Hence, $s$-wave interaction
contributes energy of order $a$ (without the correction (\ref{eq:de_ii}))
and $p$-wave interaction contributes energy of order $\mathrm{const}\cdot a^{3}k^{2}$.
However, for spinless bosons, $p$-wave interaction is forbidden (because
it is antisymmetric with respect to interchange of two bosons, see
\cite{Landau_QM}). Therefore, the correction (\ref{eq:de_ii}) presented
here for the GPE is more significant than corrections originating
from higher orders of partial waves expansion.

\selectlanguage{english}%

\section{\label{sec:ground-state-of}ground state of a thomas-fermi (TF) BEC
in a harmonic trap }

We would like to compare the ground states of the standard GPE (\ref{Time independent GP0})
and the modified GPE (\ref{eq:2Deltas GP}) in a time independent
trapping potential $U\left(x\right)$. The Thomas-Fermi (TF) approximation
\cite{Pethick@Smith} for the standard GPE (where the kinetic energy
is neglected) is\foreignlanguage{american}{
\begin{eqnarray}
U\left(x\right)\psi\left(x\right)+g\left|\psi\left(x\right)\right|^{2}\psi\left(x\right) & = & \mu\psi\left(x\right).\label{eq:thomas fermi-1}
\end{eqnarray}
Introducing the density $\rho_{0}=\left|\psi\left(x\right)\right|^{2}$,
it takes the form}

\selectlanguage{american}%
\begin{eqnarray}
\rho_{0} & =\frac{1}{g} & \left(\mu-U\left(x\right)\right)\label{eq:TF_dens}
\end{eqnarray}
where $\mu$, the chemical potential, is a constant determined by
the normalization $\int_{-R}^{R}\left|\psi\left(x\right)\right|^{2}dx=1$\foreignlanguage{english}{
and $R$ satisfies 
\begin{equation}
\mu=U\left(R\right).\label{eq:chem}
\end{equation}
The TF approximation is valid at the central region of the trap, $-R\lesssim x\lesssim R$
\cite{stringary}. For a harmonic trap} 
\begin{equation}
U\left(x\right)=\frac{1}{2}m\omega^{2}x^{2}.\label{eq:tarp}
\end{equation}
Normalization of the wave function, $\int_{-R}^{R}\left|\psi\right|^{2}dx=1$,
yields
\begin{eqnarray}
\mu & = & \left(\frac{3\sqrt{m}}{2^{5/2}}g\omega\right)^{\frac{2}{3}}\label{eq:mu}
\end{eqnarray}
and
\begin{eqnarray}
R & =\sqrt{\frac{2\mu}{m\omega^{2}}}= & \left(\frac{3}{2m}g\omega^{-2}\right)^{\frac{1}{3}}.\label{eq:R}
\end{eqnarray}
In what follows, this value of $R$ (that is independent of $l$)
will be used. The chemical potential for the standard GPE, defined
as
\begin{equation}
\mu=\int\left[-\psi^{*}\frac{\hbar^{2}}{2m}\nabla^{2}\psi+U\left(r\right)\left|\psi\right|^{2}+g\left|\psi\right|^{4}\right]dx
\end{equation}
is related to the various energy contributions\textbf{ }(kinetic energy
$E_{k}$, potential energy $E_{p}$ and non-linear energy $E_{nl})$\textbf{
}by 
\begin{equation}
\mu=E_{k}+E_{p}+2E_{nl}.\label{eq:virial}
\end{equation}
For a harmonic potential \cite[page 167]{Pita_book},
\begin{equation}
E_{k}-E_{p}+\frac{1}{2}E_{nl}=0,\label{eq:E_virial}
\end{equation}
so, if $E_{k}$ is negligible (as assumed in the TF approximation),
\begin{equation}
E_{nl}\approx2E_{p}\approx\frac{2}{5}\mu\label{eq:Es}
\end{equation}
leading to
\begin{equation}
E_{p}\approx\frac{1}{5}\mu=\frac{1}{5}\left(\frac{3\sqrt{m}}{2^{5/2}}g\omega\right)^{\frac{2}{3}}\label{eq:Epl_l0}
\end{equation}
and
\begin{eqnarray}
E_{nl} & \approx & \frac{2}{5}\mu=\frac{2}{5}\left(\frac{3\sqrt{m}}{2^{5/2}}g\omega\right)^{\frac{2}{3}}.\label{eq:Enl_l0}
\end{eqnarray}
The total energy of a particle in the condensate is given by
\begin{eqnarray}
E & = & \frac{3}{5}\mu=\frac{3}{5}\left(\frac{3\sqrt{m}}{2^{5/2}}g\omega\right)^{\frac{2}{3}}.\label{eq:gp_energy}
\end{eqnarray}

Now, let us consider the modified GPE (\ref{eq:2Deltas GP}). In the
TF approximation it takes the form

\begin{eqnarray}
U\left(x\right)\psi\left(x\right)+g\left|\psi\left(x\right)\right|^{2}\psi\left(x\right)-\frac{1}{2}gl^{2}\frac{d^{2}\left|\psi\left(x\right)\right|^{2}}{dx^{2}}\psi\left(x\right) & = & \mu\left(l\right)\psi\left(x\right)\label{eq:thomas fermi}
\end{eqnarray}
that reduces to (in analogy to (\ref{eq:TF_dens}))

\begin{eqnarray}
\left|\psi\left(x\right)\right|^{2}-\frac{1}{2}l^{2}\frac{d^{2}\left|\psi\left(x\right)\right|^{2}}{dx^{2}} & =\frac{1}{g} & \left(\mu\left(l\right)-U\left(x\right)\right)
\end{eqnarray}
\foreignlanguage{english}{and 
\begin{equation}
\rho\left(x\right)=\left|\psi\left(x\right)\right|^{2}\approx\frac{1}{g}\left(\mu\left(l\right)-U\left(x\right)\right)-\frac{1}{2g}l^{2}\frac{d^{2}U\left(x\right)}{dx^{2}}.\label{eq:TF_zinner}
\end{equation}
A similar differential equation was studied and solved in \cite{Jensen_TF}
(for discussion regarding the stability of the solutions see \cite{Jensen_stab}).
However, we assume that the term $\frac{1}{2g}l^{2}\frac{d^{2}\rho\left(x\right)}{dx^{2}}$
in (\ref{eq:TF_zinner}) can be considered as a perturbation so that
for a harmonic trap (\ref{eq:tarp}) one finds} 
\begin{eqnarray}
\rho\left(x\right) & \approx & \frac{1}{g}\left(\mu\left(l\right)-\frac{1}{2}m\omega^{2}x^{2}-\frac{1}{2}m\omega^{2}l^{2}\right).\label{eq:density}
\end{eqnarray}
This density differs from the standard GPE density 
\begin{equation}
\rho_{0}\left(x\right)=\frac{1}{g}\left(\mu\left(0\right)-\frac{1}{2}m\omega^{2}x^{2}\right)\label{eq:parabola}
\end{equation}
 by a small negative constant\foreignlanguage{english}{ 
\begin{equation}
\rho-\rho_{0}=-\Delta\rho=\frac{\Delta\mu}{g}-\frac{1}{2g}m\omega^{2}l^{2}\label{eq:droh0}
\end{equation}
where $\Delta\mu=\mu\left(l\right)-\mu\left(0\right)$.}

\selectlanguage{english}%
The TF approximation is valid only at the central part of the trap,
where the density of atoms is very large. Since both $\rho\left(x\right)$
and $\rho_{0}\left(x\right)$ are normalized to $1$, we expect that
on the edges of the condensate, where the second derivative of the
density is positive, $\rho\left(x\right)$ will be higher than $\rho_{0}\left(x\right)$.
The edge is defined by $R-2d<\left|x\right|<R$, \foreignlanguage{american}{where
$d$, the typical thickness of the boundary, satisfies \cite{stringary}
\[
\left.\frac{dU}{dr}\right|_{R}\cdot d=\frac{\hbar^{2}}{2md^{2}}
\]
or
\begin{equation}
d=\left(\frac{2m}{\hbar^{2}}\left.\frac{dU}{dx}\right|_{R}\right)^{-1/3}=\left(\frac{2m^{2}}{\hbar^{2}}\omega^{2}R\right)^{-1/3}.\label{eq:d}
\end{equation}
In Sec. IV and in App. B, $\Delta\rho$ is calculated (see Eq. (\ref{eq:101}))
and is found to take the value of}
\begin{equation}
\Delta\rho=\frac{1}{4g}m\omega^{2}l^{2}\label{eq:droh}
\end{equation}
resulting in\foreignlanguage{american}{
\begin{equation}
\Delta\mu=\frac{1}{4}m\omega^{2}l^{2}.\label{eq:mu_cor}
\end{equation}
}

It is possible to calculate both $\rho_{0}\left(x\right)$ and $\rho\left(x\right)$
numerically. Numerical determination of the ground state is generally
carried out by propagating in imaginary time, i.e. one replaces $\delta t$
with $-i\delta t$ in the split step evolution operator and normalizes
the wavefunction to one after each time step. We use the evolution
operator
\begin{equation}
\hat{P}=exp\left[-\frac{\delta t}{\hbar}\left(-\frac{\hbar^{2}}{2m}\frac{d^{2}}{dx^{2}}+U\left(x\right)+2g\left|\psi\left(x\right)\right|^{2}-\frac{1}{2}\left|\psi\left(x+l\right)\right|^{2}-\frac{1}{2}\left|\psi\left(x-l\right)\right|^{2}\right)\right],\label{eq:evo_old}
\end{equation}
corresponding to the time dependent version of (\ref{eq:2Deltas GP}),
and stop the propagation when a steady state is reached, i.e.,
\begin{equation}
\hat{P}\psi=\overline{\lambda}\psi\label{eq:lam}
\end{equation}
where the factor $\bar{\lambda}$ \textbf{(}which is close to one
for small $\delta t$\textbf{) }is eliminated after normalization. 

\begin{figure}[H]
\selectlanguage{english}%
(a) \qquad{}\qquad{}\qquad{}\qquad{}\qquad{}\qquad{}\qquad{}\qquad{}\qquad{}\qquad{}\qquad{}\qquad{}(b)

\includegraphics[clip,scale=0.4]{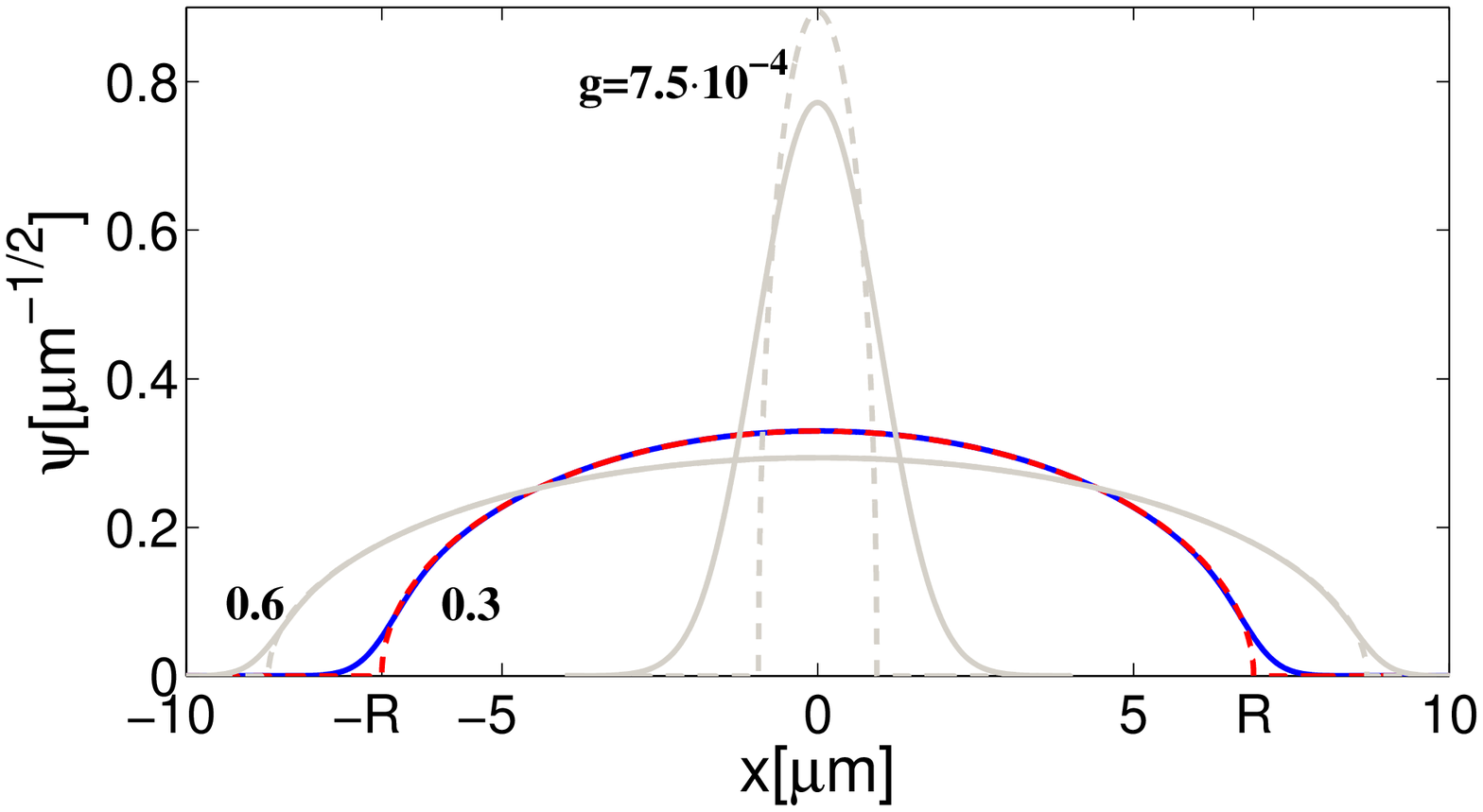} \qquad{}\qquad{}\includegraphics[clip,scale=0.4]{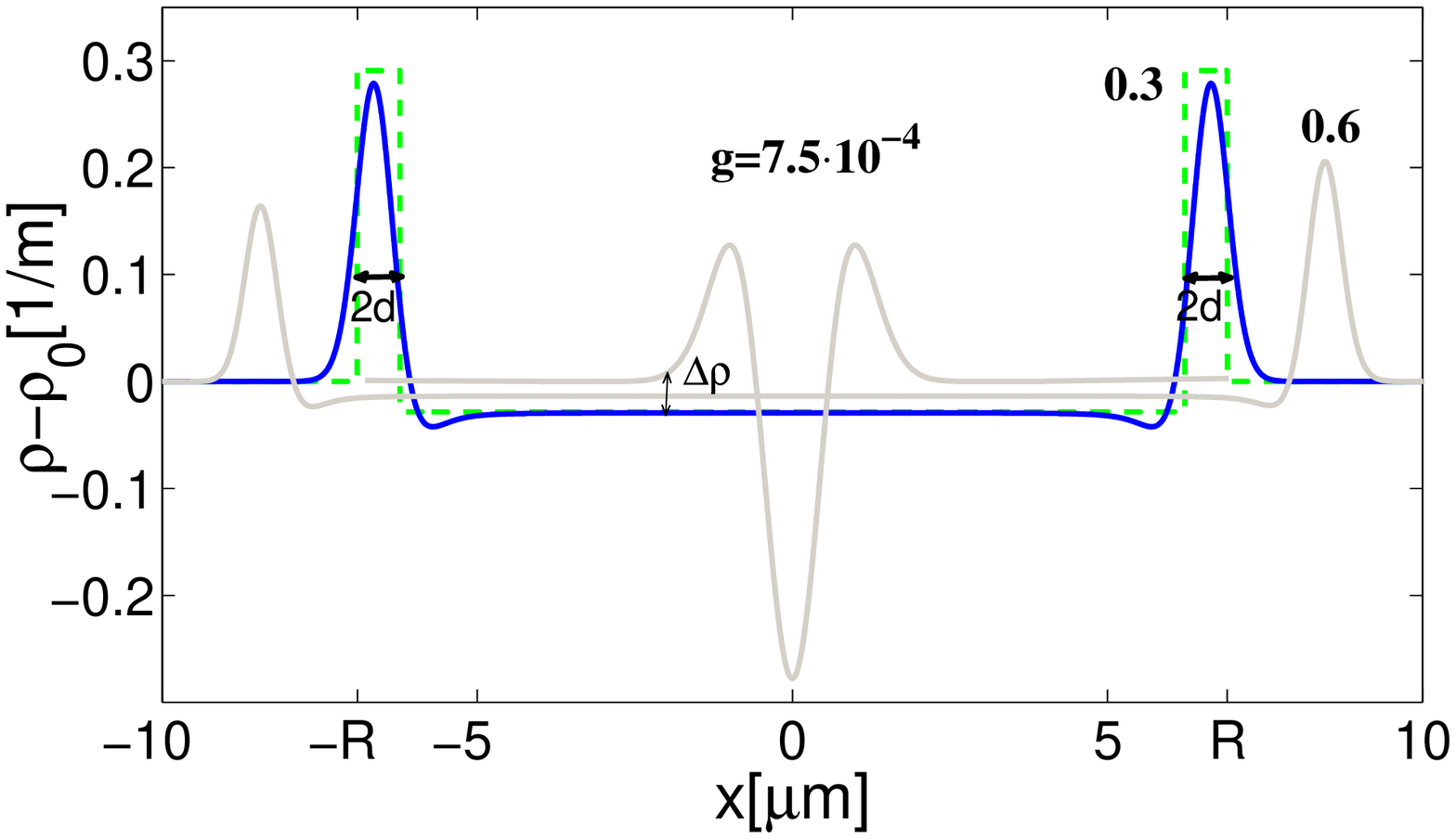}

\qquad{}\qquad{}\qquad{}\qquad{}\qquad{}(c)

\qquad{}\qquad{}\qquad{}\qquad{}\qquad{}\includegraphics[clip,scale=0.4]{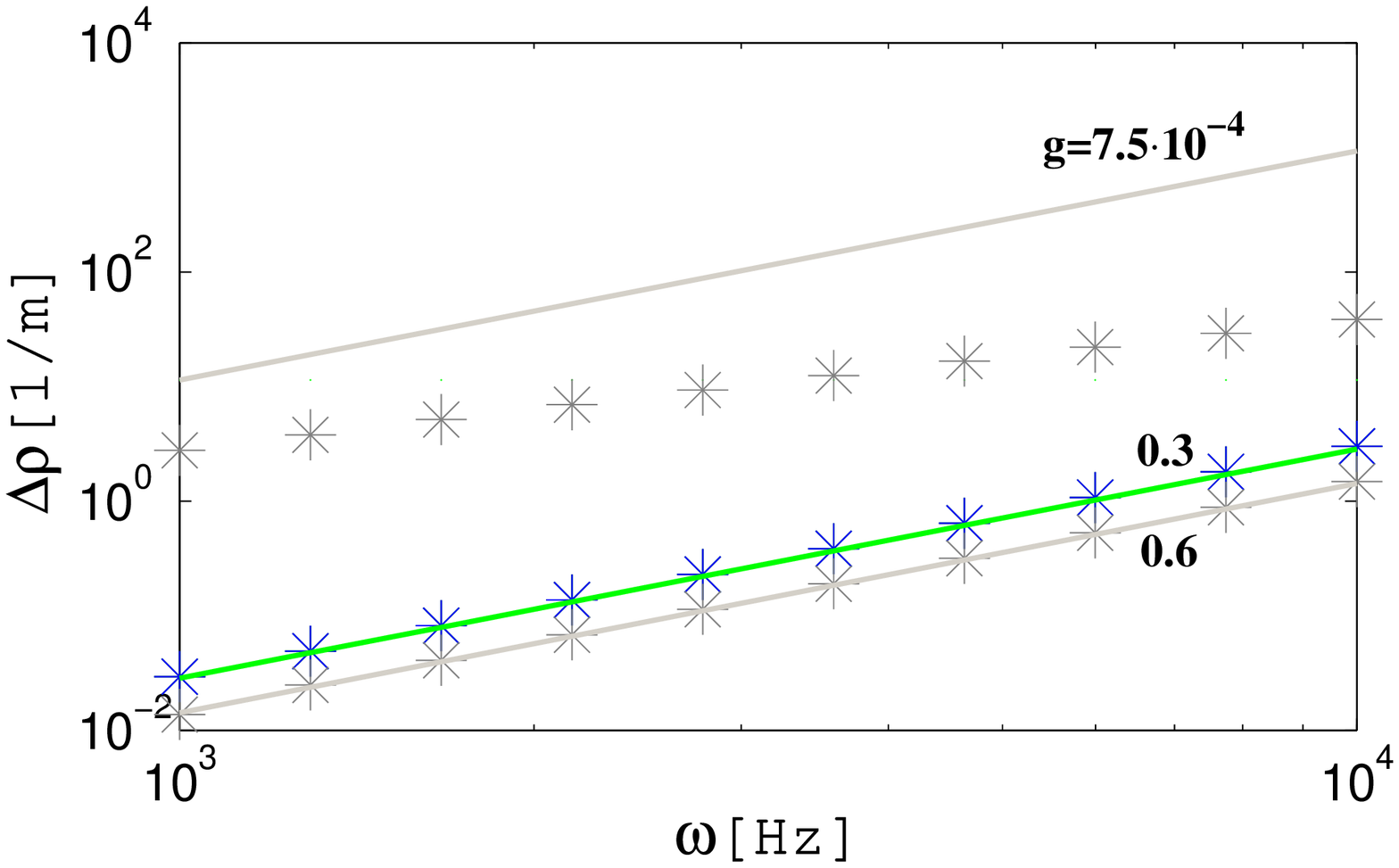}

\selectlanguage{american}%
\caption{\label{fig:wavefunction}\foreignlanguage{english}{(Color online)}
ground state of Rubidium atoms $m=87\left[\mathrm{amu}\right]$ for
various interaction parameters $g$ marked on the figure. The results
for $g=0.3\left[\mathrm{Hz\cdot m}\right]$\foreignlanguage{english}{
}are colored while the ones for stronger or weaker nonlinearities
are marked by light gray. If $g$ is too small, the TF approximation
is not valid and there is a big difference between the theory and
the numerical results. (a) Ground state wave function of the GPE calculated
with the help of (\ref{eq:evo_old}) (solid line) and the Thomas-Fermi
approximation (\ref{eq:density}) or (\ref{eq:parabola}) (dashed
line) for $\omega=1\left[\mathrm{kHz}\right]$. (b) Solid line - Calculated
density deviations (using (\ref{eq:evo_old})) due to nonzero van-der-Waals
radius $\rho\left(l=5\left[nm\right]\right)-\rho_{0}\left(l=0\right)$\textbf{,}
$\omega=1\left[\mathrm{kHz}\right]$. Dashed green line - simple approximation
(\ref{eq:simple_density}) for the deviations. Here $d=0.3\left[\mathrm{\mu m}\right]$
(for $g=0.3\left[\mathrm{Hz\cdot m}\right]$) , as can be found from
(\ref{eq:R}) and (\ref{eq:d}). Note that the deviations between
the TF results and the GP ones as well as between $\rho$ and $\rho_{0}$
are largest in a region of size $d$ around $x=R$. The deviations
for $g=7.5\cdot10^{-4}\left[\mathrm{Hz\cdot m}\right]$ are divided
by $10$ is order to make the figure clear. (c) Density deviations
at the center $x=0$ of the trap $\Delta\rho=\rho_{0}\left(l=0\right)-\rho\left(l=5\left[nm\right]\right)$
as a function of the trapping frequency $\omega$. The solid line
is the prediction (\ref{eq:droh}) and the stars are the numerically
calculated values using (\ref{eq:evo_old}). The scale is logarithmic\textbf{.}}
\end{figure}

This scheme works better if we choose the initial wavefunction to
be close to the ground state. We take the approximate ground state
\foreignlanguage{american}{(\ref{eq:parabola}) as an initial wavefunction.
Propagating the modified GPE (\ref{eq:2Deltas GP}) in a split-step
technique with imaginary time steps minimizes the energy and gives
the perturbed ground state for particles interacting with a potential
of finite range. We repeat this calculation for various values of
the interaction range $l$ (including $l=0$) and see that (\ref{eq:density})
and (\ref{eq:droh0}) are satisfied in the central region of the trap
(Fig. \ref{fig:wavefunction}) with (\ref{eq:droh}) and (\ref{eq:mu_cor}).}One
should distinguish between the correction to the TF approximation
in the vicinity of $x=R$ (Fig. 1a) and the correction resulting of
the non-vanishing value of $l$ (Figs. 1b and c). Note that our analytical
results are valid \textbf{only }when the TF approximation holds.

\selectlanguage{american}%

\section{\label{sec:energy-of-a}energy of a thomas-fermi bec in harmonic
trap}

The energy of a BEC according to the modified GPE (\ref{eq:2Deltas GP})\foreignlanguage{english}{
is a sum of kinetic, potential and nonlinear contributions. In the
Thomas-Fermi (TF) approximation, we neglect the kinetic energy and
we are left with potential energy
\begin{equation}
E_{p}=\int\psi^{*}\left(x\right)U\left(x\right)\psi\left(x\right)dx=\int\rho\left(x\right)U\left(x\right)dx\label{eq:Epd}
\end{equation}
and with interaction energy}
\begin{eqnarray}
E_{nl} & = & \frac{g}{2}\int dx\left[\left|\psi\left(x\right)\right|^{4}-\frac{1}{2}l^{2}\frac{d^{2}\left|\psi\left(x\right)\right|^{2}}{dx^{2}}\left|\psi\left(x\right)\right|^{2}\right]\label{eq:enld}\\
 & = & \frac{g}{2}\int dx\left[\rho^{2}\left(x\right)-\frac{1}{2}l^{2}\frac{d^{2}\rho\left(x\right)}{dx^{2}}\rho\left(x\right)\right].\nonumber 
\end{eqnarray}
In the ground state of the harmonic trap with only contact interaction
$\left(l=0\right)$, these energies are given by (\ref{eq:Epl_l0})
and (\ref{eq:Enl_l0}). 

\begin{figure}[H]
\selectlanguage{english}%
(a)

\includegraphics[scale=0.4]{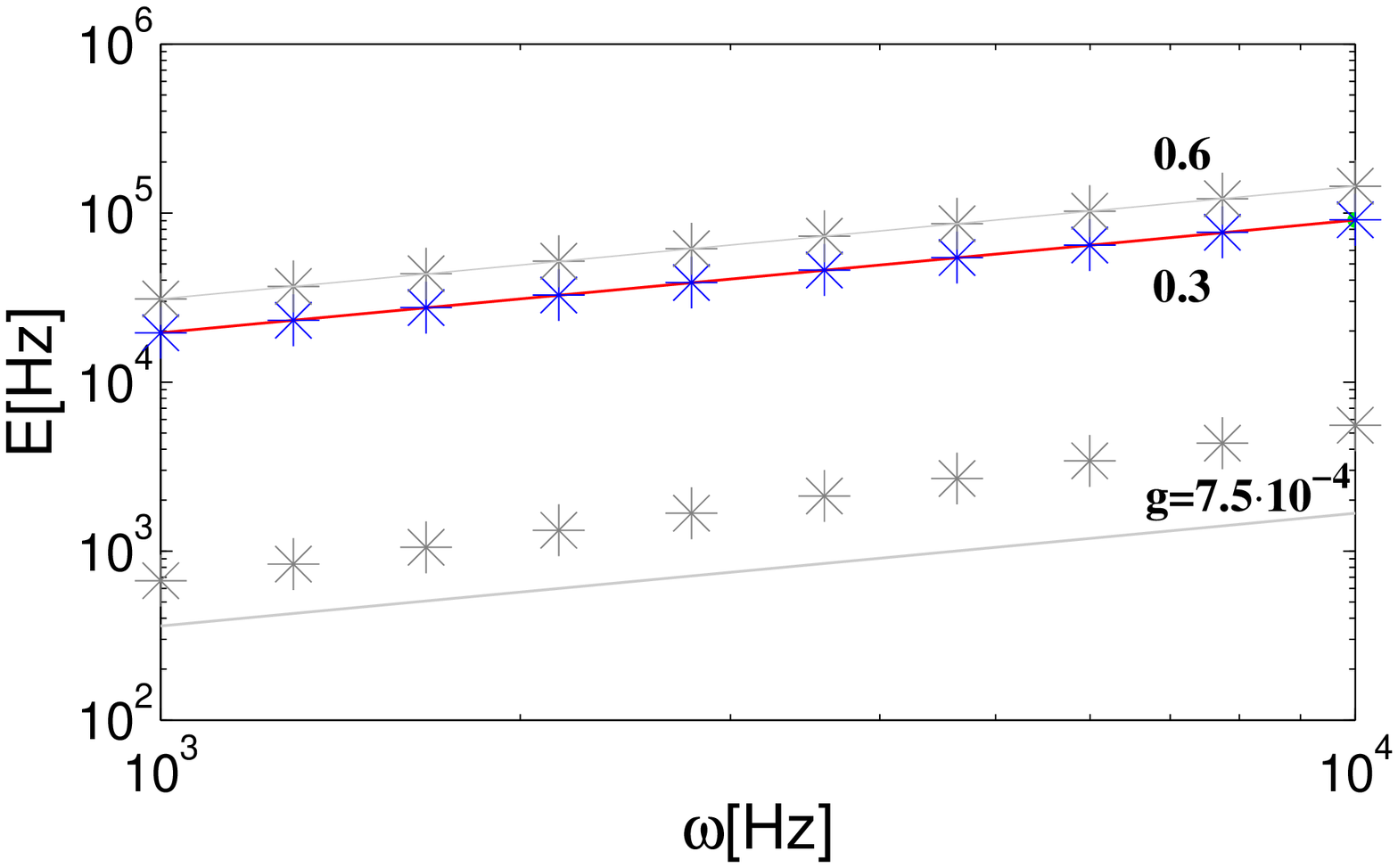}

(b)

\includegraphics[scale=0.4]{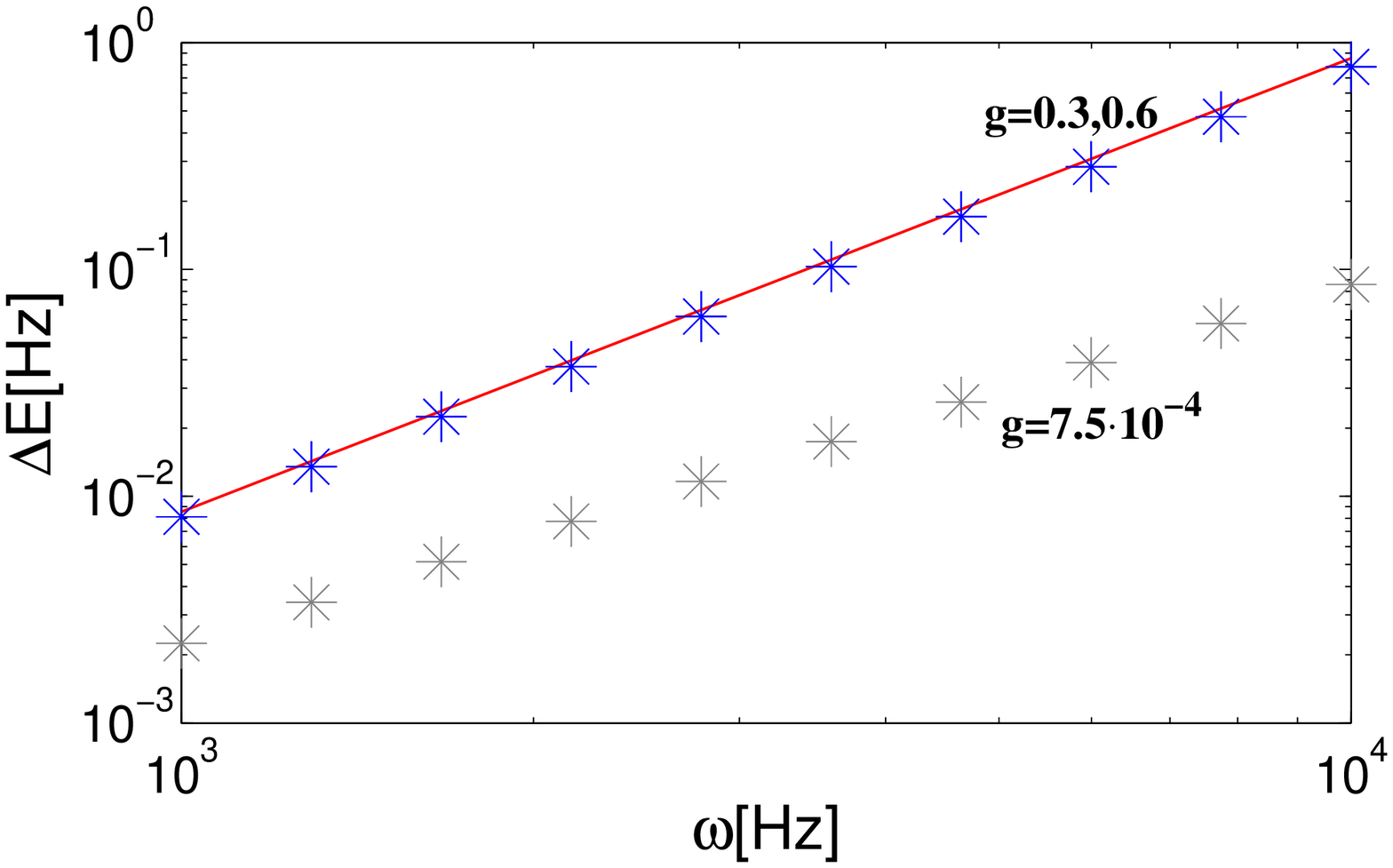}

\selectlanguage{american}%
\caption{\selectlanguage{english}%
\label{fig:ground_Energy}(Color online) Energy of the ground state
\foreignlanguage{american}{of Rubidium BEC }as a function of the trap
frequency $\omega$\foreignlanguage{american}{ for various interaction
parameters $g$ marked on the figure. The results for $g=0.3\left[\mathrm{Hz\cdot m}\right]$}
\foreignlanguage{american}{are colored while the ones for stronger
or weaker nonlinearities are marked by light gray.} (a) Energy for
$l=0$.solid line - the prediction (\ref{eq:gp_energy}), stars -
the calculated values integrating (\ref{eq:evo_old}). (b) The correction
to the energy $\Delta E\left(l\right)=E\left(l=5\left[nm\right]\right)-E\left(l=0\right)$.
solid red line - the prediction (\ref{eq:tot_corE}), stars - the
calculated values integrating (\ref{eq:evo_old}).\selectlanguage{american}%
}
\end{figure}

The ground state energy of the modified GPE (\ref{eq:2Deltas GP})
\begin{equation}
E=E_{p}+E_{nl}
\end{equation}
can be written in the form
\begin{equation}
E=E\left(l=0\right)+\Delta E\left(l\right).
\end{equation}
Assuming that in the regime where the TF approximation holds the deviation
of $\rho$ from $\rho_{0}$ is a constant denoted by $\delta\rho$,
$E\left(l=0\right)$ can be considered as a minimum of 
\begin{equation}
E_{0}=E\left(l=0\right)+C_{2}^{GP}\left(\delta\rho\right)^{2}
\end{equation}
with respect to $\delta\rho$, with the constant $C_{2}^{GP}>0$ (see
App. B, (\ref{eq:49b})). The ground state energy of the modified
GPE is the minimum of 
\begin{equation}
E=E\left(l=0\right)+C_{2}^{GP}\left(\delta\rho\right)^{2}+C_{0}^{pert}\left(l\right)+C_{1}^{pert}\left(l\right)\delta\rho\label{eq:**}
\end{equation}
where we expand $\Delta E$ in powers of $\delta\rho$ with constants
$C_{0}^{pert}$ and $C_{1}^{pert}$ (see (\ref{eq:C0_ex}) and (\ref{eq:C1_ex})
in App. B). This minimum is obtained for
\begin{equation}
\delta\rho=\Delta\rho=-\frac{C_{1}^{pert}}{2C_{2}^{GP}}\label{eq:d_rho_cal}
\end{equation}
and the resulting value of $E$ is
\begin{equation}
E=E\left(l=0\right)-\frac{\left(C_{1}^{pert}\right)^{2}}{2C_{2}^{GP}}+C_{0}^{pert}\left(l\right).
\end{equation}
The parameters $C_{2}^{GP},\, C_{1}^{pert}$ and $C_{0}^{pert}\left(l\right)$
are calculated explicitly in App. B (Eqs. (\ref{eq:49b}), (\ref{eq:C1_ex})
and (\ref{eq:C0_ex})). From (\ref{eq:enld}) we see that $C_{0}^{pert}\left(l\right)$
and \foreignlanguage{english}{$C_{1}^{pert}\left(l\right)$} are proportional
to $l^{2}$. Therefore in the leading order $\frac{C_{1}^{pert}\left(l\right)^{2}}{2C_{2}^{GP}}$
can be neglected. The leading order correction to the energy which
is related to the van-der-Waals radius is (see (\ref{eq:C0_ex}))
\begin{equation}
\Delta E_{0}\left(l\right)\approx C_{0}^{pert}=-\frac{g}{4}\int_{-R}^{R}l^{2}\frac{d^{2}\rho_{0}\left(x\right)}{dr^{2}}\rho_{0}\left(x\right)dx.
\end{equation}
and using (\ref{eq:parabola}) we obtain in the leading order in $l^{2}$
\begin{equation}
\Delta E\left(l\right)\approx\frac{1}{4}ml^{2}\omega^{2}.\label{eq:tot_corE}
\end{equation}
This is the main result of the present work. Since the correction
to the chemical potential given by (\ref{eq:mu_cor}) turns out not
to depend on the density, we obtain the same correction for the energy
per particle (\ref{eq:tot_corE}). The correction (\ref{eq:tot_corE})
is very small compared to the total energy (\ref{eq:gp_energy}),\textbf{
}
\begin{equation}
\frac{\Delta E\left(l\right)}{E\left(l=0\right)}=\frac{5l^{2}}{12}\left(\frac{2^{5/2}m\omega^{2}}{3g}\right)^{\frac{2}{3}}=\frac{5}{6}\cdot\frac{l^{2}}{R^{2}}=-\frac{5}{3}\frac{g_{2}}{R^{2}}.\label{eq:ratio}
\end{equation}
where $R$ is given by (\ref{eq:R}). Although the correction (\ref{eq:tot_corE})
is small for realistic parameters, we believe that it can be measured
because it is linear in $\omega^{2}$ while $E\left(l=0\right)\propto\omega^{2/3}$
(\ref{eq:gp_energy}) (see Fig \ref{fig:ground_Energy}). In the discussion
(Sec. V), we present estimates for the magnitude of the correction
(\ref{eq:tot_corE}). In particular, a possibility to substantially
increase $l$ with the help of Feshbach resonances is discussed. Furthermore,
using molecules \cite{molecules,molecules_rev} or Rydberg atoms \cite{Rydberg_BEC}
instead of atoms in their ground state is likely to increase significantly
the length $l$ and hence to increases $\Delta E\left(l\right)$.
From Fig. 2(b), it is seen that in the TF regime, the correction to
the energy does not depend on $g$ (and therefore the lines for $g=0.3\left[\mathrm{Hz\cdot m}\right]$
and $g=0.6\left[\mathrm{Hz\cdot m}\right]$ merge), while for weaker
nonlinearity parameters the correction does depend on $g$ and disagrees
with our theoretical results.

\section{\label{sec:summary-and-conclusions}summary and discussion}

In this work we introduce a simple model (\ref{eq:delta shell}) taking
into account the extension of inter-particle potential, and calculate
the corrections to the standard GPE where $\delta$-function interaction
potentials potentials are assumed. For realistic experimental parameters
we find that these corrections are indeed small. We calculate the
correction for a Bose-Einstein condensate in a harmonic trap, in a
situation where in most of the volume of the trap the Thomas Fermi
(TF) approximation is valid. The correction to the ground state energy
is given in the leading order by (\ref{eq:tot_corE}), namely, 
\begin{equation}
\Delta E\left(l\right)=\frac{1}{4}ml^{2}\omega^{2}\label{eq:53}
\end{equation}
where $m$ is the mass of the particles, $l$ is the extension of
the potential (related to the van-der-Waals radius) and $\omega$
is the frequency of the confining harmonic trap. By setting $l$ equal
to the typical range of van-der Waals potentials (about $100a_{0}$
where $a_{0}=0.5\overset{\circ}{\mathrm{A}}$ is the Bohr radius)
and using a small condensate of size of $R=10\mu m\approx2\cdot10^{5}a_{0}$,
we find a relative correction (Eq. (\ref{eq:ratio})) of $10^{-6}$
which is extremely small and hard to measure. However, as discussed
in Ref. \cite{Flambaum} (see also \cite{Massignan_castin,stoof}),
the effective range diverges near Feshbach resonances and zero crossings
of the scattering length. In particular, near zero crossing \cite{Jensen_stab},
\begin{equation}
g_{2}\thicksim-\frac{r_{e0}}{2}\frac{a_{bg}^{2}}{a},
\end{equation}
where the relevant length scales are $r_{e0}$ (the effective range
at the Feshbach resonance), $a_{bg}$ (the background scattering length)
and $a$ (the scattering length). For a broad resonance, the effective
range is larger than the van-der Waals radius, for a narrow resonance,
it can be much larger \cite{van_der_walls}. Assuming both $r_{e0}$
and $a_{bg}$ are of the order of the van-der Waals radius, namely
$100a_{0}$, we estimate
\begin{equation}
g_{2}\approx-10^{6}\frac{a_{0}^{3}}{2a}.
\end{equation}
Zero crossings of scattering lengths have been resolved to within
$0.01a_{0}$ \cite{Roati}. Assuming $a\approx0.1a_{0}$, we obtain
a $g_{2}$ of $\approx-10^{7}a_{0}^{2}$. Taking $R=10\left[\mu m\right]$implies
a relative energy correction (\ref{eq:ratio}) due to the finite extension
of the potential of the order of $\frac{1}{2}\cdot$$10^{-3}$, which
could be even larger for smaller condensates (using for example atom
chips with tight confinement) and atoms with larger background scattering
lengths or narrow Feshbach resonances. The TF approximation is still
valid since for sufficiently high atom number $N$, the TF radius
$R$ (Eq. (\ref{eq:R}))
\begin{eqnarray}
R & = & \left(\frac{3}{2m}g\omega^{-2}\right)^{\frac{1}{3}}=a_{\parallel}\left(\frac{3Naa_{\parallel}}{a_{\perp}^{2}}\right)^{\frac{1}{3}}
\end{eqnarray}
can always be made larger than the oscillator length $a_{\parallel}$.
Here $a_{\parallel}=\sqrt{\frac{\hbar}{m\omega}}$ and $a_{\perp}=\sqrt{\frac{\hbar}{m\omega_{\perp}}}$
are the harmonic trap length-scales in the parallel ($x$) and the
transverse ($y,z$) directions respectively. For the scattering length
of $0.1a_{0}$ assumed above ,$a_{\parallel}$ of the order of $2\left[\mu m\right]$
and $a_{\perp}\approx0.2\left[\mu m\right]$ , an atom number of $N>10^{5}$
ensures the validity of the TF approximation.

Our results apply also to novel condensates of molecules \cite{molecules,molecules_rev,Raizen_cooling},
photons \cite{photonsBEC}, and polaritons \cite{Jaqline2014,Glazov}
where the correction to the GPE may be larger. There are other corrections
to the GPE, for example, the Lee-Yang-Huang (LYH) correction \cite{Pita_book}
which is typically larger than the correction that was considered
here and is fundamentally of a different origin, as it depends on
the density of particles while the correction (\ref{eq:53}) does
not depend on this quantity. These corrections scale differently with
the trap frequency $\omega$. In elongated traps, the LYH correction
is linear in $\omega$ and can in principle be distinguished from
the correction calculated here which is proportional to $\omega^{2}$.
Our calculations are in one dimension, but the extension to higher
dimensions is straightforward. 

The approach of capturing the physics of realistic potentials by several
delta functions should have many applications beyond the purpose of
this paper.

\section*{acknowledgments}

We appreciate useful discussions with Ofir Alon, Alain Aspect, Rob
Ballagh, Immanuel Bloch, Jacqueline Bloch, Eugene Bogomolny, Simon
Gardiner, Tal Goren, Peter Littlewood, Nimrod Moiseyev, Christopher
Pethick, Lev Pitaevsii, Jeff Steinhauer, Sandro Stringari and Nikolaj
Thomas Zinner. This work was partly supported by the National Science
Foundation, by the Israel Science Foundation (ISF) grant number 1028/12,
by the US-Israel Binational Science Foundation (BSF) grant number
2010132, by the Minerva Center of Nonlinear Physics of Complex Systems
and by the Shlomo Kaplansky academic chair.

\section*{Appendix A}

In this appendix, we calculate the phase shift for the ``$\delta$-atom''
(\ref{eq:delta shell}), which can be used for defining the one dimensional
toy model (\ref{eq:3_potential}) and determining $\frac{1}{2}l^{2}$
in (\ref{eq:2Deltas GP}).

The characteristic length scales of the interactions are small compared
to the length scales of the trapping potential and therefore we are
allowed to assume that the external potential is constant over the
interaction regime, i.e., the wavefunction $\psi$ depends only on
the relative distance $r$ between two particles and out of the pair
interaction range, the particles are assumed to be free. In the absence
of interactions, the wave function is a free wave
\begin{equation}
\psi\left(r\right)=A\frac{\sin\left(kr\right)}{r}
\end{equation}
where $k=\sqrt{2mE/\hbar^{2}}$. Interactions will add a phase shift
$\delta$, so that for $r>r_{out}$, 
\begin{equation}
\psi\left(r\right)=A\frac{\sin\left(kr+\delta\right)}{r}.\label{eq:wave}
\end{equation}
We turn now to calculate $\delta$, the phase shift caused by the
potential (\ref{eq:delta shell}). In \cite{Pethick_paper}, it is
argued that the phase shift is related to the pair interaction energy
$E_{pair}$ (Eq.6 of \cite{Pethick_paper}) by 
\begin{equation}
\begin{array}{ccc}
E_{pair} & = & \frac{2\hbar^{2}k}{m}\left(-\frac{\delta}{L}\right)\end{array}.\label{eq:dE_pet}
\end{equation}
Here, the non-linear energy of each particle is 
\begin{equation}
E_{nl}=N\cdot E_{pair}
\end{equation}
where $E_{nl}$ is the change in the energy that the scattering potential
could cause if the particles were constrained to stay in a large ball
of radius\textbf{ $L$ }while $E_{pair}$ is the interaction energy
of a single pair of particles.\textbf{ }The energy shift $E_{pair}$
is calculated using perturbation theory of the lowest order in $r_{in}$
, $\lambda$ and $g_{3D}$: 
\begin{equation}
\begin{array}{ccl}
E_{pair} & \approx & \int_{0}^{r_{in}}4\pi r^{2}\cdot\left|\psi\left(r\right)\right|^{2}\cdot\frac{3\left(g_{3D}+\lambda\right)}{4\pi r_{in}^{3}}dr-\int_{0}^{\infty}4\pi r^{2}\cdot\left|\psi\left(r\right)\right|^{2}\cdot\frac{\lambda}{4\pi r_{out}^{2}}\delta\left(r-r_{out}\right)dr\\
 & = & \frac{3\left(g_{3D}+\lambda\right)}{r_{in}^{3}}\int_{0}^{r_{in}}r^{2}\cdot\left|\psi\left(r\right)\right|^{2}dr-\lambda\cdot\left|\psi\left(r_{out}\right)\right|^{2}.
\end{array}\label{eq:de}
\end{equation}
Using $\psi\left(r\right)$ of (\ref{eq:wave}), and in the limit
$r_{in}\rightarrow0$ one finds 
\begin{equation}
E_{pair}\approx A^{2}\left[\left(g_{3D}+\lambda\right)\cdot k^{2}-\lambda\cdot\frac{\sin^{2}\left(kr_{out}\right)}{r_{out}^{2}}\right].\label{eq:E_pair}
\end{equation}
The normalization constant $A$ should satisfy 
\begin{equation}
A^{2}\int_{0}^{L}4\pi\sin^{2}\left(kr\right)dr=1.\label{A}
\end{equation}
Remembering that $L$ is very large,
\begin{equation}
\int_{0}^{L}\sin^{2}\left(kr\right)dr=\frac{1}{2}\int_{0}^{L}\left(1-\cos\left(2kr\right)\right)dr=\frac{L}{2}-\frac{1}{2k}\sin\left(2kL\right)\approx\frac{L}{2}
\end{equation}
and therefore $A=1/\sqrt{2\pi L}.$ Combining (\ref{eq:dE_pet}) and
(\ref{eq:E_pair}), we end up with
\begin{equation}
\begin{array}{ccl}
\delta & = & -\frac{LmE_{pair}}{2\hbar^{2}k}\\
 & = & -\frac{m}{4\pi\hbar^{2}k}\left[\left(g_{3D}+\lambda\right)\cdot k^{2}-\lambda\cdot\frac{\sin^{2}\left(kr_{out}\right)}{r_{out}^{2}}\right]\\
 & \approx & \frac{m}{4\pi\hbar^{2}}\left[-\left(g_{3D}+\lambda\right)k+\lambda\left(k-\frac{2}{3}r_{out}^{2}k^{3}\right)\right]
\end{array}\label{eq:d1}
\end{equation}
Here we used the fact that the wavelength is large. This leads to
a total phase shift of
\begin{equation}
\delta=-\frac{g_{3D}m}{4\pi\hbar^{2}}k-\frac{m}{4\pi\hbar^{2}}\cdot\frac{2}{3}r_{out}^{2}\cdot\lambda k^{3}.
\end{equation}
In order to calculate the scattering length and the effective range,
we write $k\cot\left(\delta\right)$ as a power series in $k$:
\begin{equation}
k\cot\left(\delta\right)=-\frac{1}{a}+\frac{1}{2}r_{e}k^{2}.\label{eq:kcot}
\end{equation}
Define $C_{3}\equiv-\frac{m\cdot\lambda}{4\pi\hbar^{2}}\cdot\frac{2}{3}r_{out}^{2}$
so that $\delta=-ak+C_{3}k^{3}$ and
\begin{equation}
\begin{array}{ccl}
k\cot\left(\delta\right) & = & \frac{k}{\tan\left(\delta\right)}\\
 & \approx & \frac{k}{\delta}\left(1-\frac{1}{3}\delta^{2}\right)
\end{array}
\end{equation}
where it is assumed that $\delta\ll2\pi$. The expansion in a power
series of $k$ yields
\begin{equation}
\begin{array}{ccl}
k\cot\left(\delta\right) & \approx & \frac{1}{-a+C_{3}k^{2}}\left(1-\frac{1}{3}\left(-a+C_{3}k^{2}\right)^{2}k^{2}\right)\\
 & \approx & -\frac{1}{a}\left(1+\frac{C_{3}}{a}k^{2}\right)\left(1-\frac{1}{3}a^{2}k^{2}\right)\\
 & \approx & -\frac{1}{a}\left[1+\left(\frac{C_{3}}{a}-\frac{1}{3}a^{2}\right)k^{2}\right].
\end{array}\label{eq:C1C3}
\end{equation}
According to (\ref{eq:kcot}) and (\ref{eq:C1C3}),
\begin{equation}
r_{e}=-2\left[\frac{C_{3}}{a^{2}}-\frac{1}{3}a\right]
\end{equation}
so that
\begin{equation}
g_{2}=a\left(\frac{a}{3}-\frac{r_{e}}{2}\right)=\frac{C_{3}}{a}=-\frac{2\lambda}{3g_{3D}}r_{out}^{2},\label{eq:g23D}
\end{equation}
resulting in the identification of $\Delta E_{int}$ of (\ref{eq:de_ii}).

\section*{Appendix B}

In this Appendix we calculate explicitly quantities used in sections
\foreignlanguage{english}{III and IV.} It turns out that in the leading
order, the correction to the energy (\ref{eq:tot_corE}) does not
depend on neither $\Delta\rho$ nor the coefficients $C_{2}^{GP}$
and $C_{1}^{GP}$ (to be defined in (\ref{eq:dnlGP12})). However,
we would like to compute it and find the analytical justification
for (\ref{eq:droh}). Using a variational principle, we analytically
calculate $\Delta\rho$ which is the value of $\delta\rho$ which
minimizes (\ref{eq:**}) and obtain the result (\ref{eq:droh}). The
expansion (\ref{eq:**}) of the energy as a power series in $\delta\rho$
takes into account the energy corrections $\Delta E_{p}=E_{p}-E_{p}\left(l=0\right)$
and $\Delta E_{nl}=E_{nl}-E_{nl}\left(l=0\right)$, given by 
\begin{eqnarray}
\Delta E_{p} & = & \int\left(\rho\left(x\right)-\rho_{0}\left(x\right)\right)U\left(x\right)dx\label{eq:pot_cor}
\end{eqnarray}
and
\begin{equation}
\Delta E_{nl}=\frac{g}{2}\int dx\left\{ \left(\rho^{2}\left(x\right)-\rho_{0}^{2}\left(x\right)\right)-\frac{1}{2}l^{2}\frac{d^{2}\rho\left(x\right)}{dx^{2}}\rho\left(x\right)\right\} .\label{eq:nl_cor}
\end{equation}
In the regime where the Thomas-Fermi (TF) approximation is valid,
$-R\lesssim x\lesssim R$, we showed (Eq. (\ref{eq:droh0})) that
for a harmonic potential, $\rho\left(x\right)=\rho_{0}\left(x\right)-\delta\rho$,
where $\delta\rho$ is small and does not depend on $x$. At $x\approx\pm R$,
$\rho\left(x\right)-\rho_{0}\left(x\right)$ has sharp picks with
total integrated area of approximately $2R\delta\rho$. Thus, it is
convenient to introduce a simplified density
\begin{equation}
\rho\left(x\right)=\begin{cases}
\rho_{0}\left(x\right)-\delta\rho & \left|x\right|<R-2d\\
\rho_{0}\left(x\right)+\frac{R}{2d}\delta\rho\:\quad & R-2d<\left|x\right|<R
\end{cases}\label{eq:simple_density}
\end{equation}
for calculating the energies. This density (dashed line in Fig \ref{fig:wavefunction}b)
assumes that the correction resulting of the finite range of interaction
is piecewise constant. It can be used to estimate integrals involving
the density and smooth quantities. This approximate density was introduced
since we know to calculate the density only in $\left[0,R-2d\right]$
where the TF approximation is valid. The estimate (\ref{eq:simple_density})
of $\rho\left(x\right)$ in the interval $\left[R-2d,R\right]$ relies
on the fact that both $\rho$ and $\rho_{0}$ are normalized to $1$.
Using the density (\ref{eq:simple_density}) and the relation (\ref{eq:R})
for calculating the deviation (\ref{eq:pot_cor}), one finds (taking
for each order of $\delta\rho$ only the leading term in $\frac{d}{R}$,
assumed to be small when the TF approximation is valid), 
\begin{equation}
\begin{array}{ccl}
\Delta E_{p} & \approx & \delta\rho\left[-\int_{-R+2d}^{R-2d}U\left(x\right)dr+\frac{R}{d}\int_{R-2d}^{R}U\left(x\right)dr\right]\\
 & = & \delta\rho\left\{ -\frac{1}{3}m\left(R-2d\right)^{3}\omega^{2}+m\frac{R}{6d}\left[R^{3}-\left(R-2d\right)^{3}\right]\omega^{2}\right\} \\
 & = & mR^{3}\omega^{2}\delta\rho\left(\frac{2}{3}-\frac{8}{3}\frac{d^{2}}{R^{2}}+\frac{8}{3}\frac{d^{3}}{R^{3}}\right)\\
 & \approx & \frac{2}{3}mR^{3}\omega^{2}\delta\rho=g\delta\rho.
\end{array}\label{eq:cor_Ep}
\end{equation}
The deviation (\ref{eq:nl_cor}) can be divided in two parts 
\begin{equation}
\Delta E_{nl}=\Delta E_{nl}^{GP}+\Delta E_{nl}^{pert}.
\end{equation}
The first contribution to $\Delta E_{nl}$, caused only by the changes
in the wavefunction, is

\begin{eqnarray}
\Delta E_{nl}^{GP} & = & \frac{g}{2}\int dx\left(\rho^{2}\left(x\right)-\rho_{0}^{2}\left(x\right)\right)\nonumber \\
 & \approx & g\int_{-R+2d}^{R-2d}\left[-\rho_{0}\left(x\right)\delta\rho+\frac{1}{2}\delta\rho^{2}\right]dx+2g\int_{R-2d}^{R}\left[\rho_{0}\left(x\right)\frac{R}{2d}\delta\rho+\frac{R^{2}}{8d^{2}}\delta\rho^{2}\right]dx\label{eq:eq:E_nl1}
\end{eqnarray}
that can be written in the form
\begin{equation}
\Delta E_{nl}^{GP}=C_{1}^{GP}\delta\rho+C_{2}^{GP}\delta\rho^{2},\label{eq:dnlGP12}
\end{equation}
and the second contribution, caused by the additional term in (\ref{eq:nl_cor}),
is
\begin{eqnarray}
\Delta E_{nl}^{pert} & = & -\frac{g}{4}\int l^{2}\frac{d^{2}\rho\left(x\right)}{dx^{2}}\rho\left(x\right)dx\nonumber \\
 & \approx & -\frac{g}{4}\int_{-R+2d}^{R-2d}l^{2}\frac{d^{2}\rho_{0}\left(x\right)}{dx^{2}}\left[\rho_{0}\left(x\right)-\delta\rho\right]dx\nonumber \\
 &  & -\frac{g}{2}\int_{R-2d}^{R}l^{2}\frac{d^{2}\rho_{0}\left(x\right)}{dx^{2}}\left[\rho_{0}\left(x\right)+\frac{R}{2d}\delta\rho\right]dx.\label{eq:E_nl_ex}
\end{eqnarray}
that can be written in the form
\begin{equation}
\Delta E_{nl}^{pert}=C_{0}^{pert}+C_{1}^{pert}\delta\rho.\label{eq:dEnl_pert}
\end{equation}
 Remembering that $\int_{-R+2d}^{R-2d}\rho_{0}\left(x\right)\approx1$
, we get
\begin{equation}
C_{1}^{GP}=-g+\frac{Rg}{d}\int_{R-2d}^{R}\rho_{0}\left(x\right)dx\label{eq:49a}
\end{equation}
\begin{equation}
C_{2}^{GP}=g\left(R-2d\right)+g\frac{R^{2}}{2d}\approx g\frac{R^{2}}{2d}\label{eq:49b}
\end{equation}
\begin{equation}
C_{0}^{pert}=\frac{1}{4}ml^{2}\omega^{2}-\frac{g}{2}\int_{R-2d}^{R}l^{2}\rho_{0}\left(x\right)\frac{d^{2}\rho_{0}\left(x\right)}{dx^{2}}dx\label{eq:49c}
\end{equation}
\begin{equation}
\begin{array}{ccl}
C_{1}^{pert} & = & -\frac{1}{4}ml^{2}\omega^{2}\left(2R-4d\right)-\frac{gR}{4d}l^{2}\int_{R-2d}^{R}\frac{d^{2}\rho_{0}\left(x\right)}{dx^{2}}dx\end{array}\label{eq:49d}
\end{equation}
We turn now to calculate the various terms assuming $\frac{d}{R}\ll1$
and begin by estimating the second term in (\ref{eq:49a}). Since
$\rho_{0}\left(x\right)$ decreases with $x$ for $x>0$, 
\begin{equation}
\frac{Rg}{d}\int_{R-2d}^{R}\rho_{0}\left(x\right)dx\ll\frac{Rg}{d}\cdot2d\cdot\rho_{0}\left(R-2d\right)=2Rg\cdot\rho_{0}\left(R-2d\right)
\end{equation}
For $\rho_{0}\left(R-2d\right)$ we can use the TF approximation (\ref{eq:parabola})
and (\ref{eq:chem}):
\begin{equation}
2Rg\cdot\rho_{0}\left(R-2d\right)\approx2R\left(\mu-\frac{1}{2}m\omega^{2}\left(R^{2}-4Rd\right)\right)=4R^{2}m\omega^{2}d=8\mu d\label{eq:69}
\end{equation}
Using (\ref{eq:mu}) and (\ref{eq:R}) we see that 
\begin{equation}
\mu R=\frac{3}{4}g\label{eq:70}
\end{equation}
 and therefore $8\mu d=6g\cdot\frac{d}{R}$ is much smaller than $g$
taking into account $d\ll R$.

Hence, it is justified to estimate
\begin{equation}
C_{1}^{GP}\approx-g\label{eq:c1GP}
\end{equation}
as expected (since $\delta\rho$ minimizes the energy for $l=0$,
the sum of $\Delta E_{p}$ and $C_{1}^{Gp}\delta\rho$ must vanish).
Now we turn to estimate $C_{1}^{pert}$. In (\ref{eq:49d}), we have
a term proportional to $\int_{R-2d}^{R}\frac{d^{2}\rho_{0}\left(x\right)}{dx^{2}}dx$.
We do not have an explicit expression for the function $\rho_{0}$
in the interval $\left[R-2d,R\right]$, but since $\frac{d\rho_{0}}{dx}\approx0$
for $x>R$, 
\begin{equation}
0=\int_{0}^{R-2d}\frac{d^{2}\rho_{0}\left(x\right)}{dx^{2}}+\int_{R-2d}^{R}\frac{d^{2}\rho_{0}\left(x\right)}{dx^{2}}dx.\label{eq:57}
\end{equation}
In the interval $\left[0,R-2d\right]$ the TF approximation results
in (\ref{eq:parabola}), so,
\begin{equation}
\int_{R-2d}^{R}\frac{d^{2}\rho_{0}\left(x\right)}{dx^{2}}dx=-\int_{0}^{R-2d}\frac{d^{2}\rho_{0}\left(x\right)}{dx^{2}}dx\approx\frac{\left(R-2d\right)m\omega^{2}}{g}.\label{eq:second_integral}
\end{equation}
The coefficient $C_{1}^{pert}$ (\ref{eq:49d}) is given by
\begin{equation}
\begin{array}{ccl}
C_{1}^{pert} & = & -\frac{1}{4}ml^{2}\omega^{2}\left(2R-4d\right)-\frac{gR}{4d}l^{2}\int_{R-2d}^{R}\frac{d^{2}\rho_{0}\left(x\right)}{dx^{2}}dx\\
 & \approx & -\frac{1}{4}ml^{2}\omega^{2}\left(2R-4d\right)-\frac{R}{4d}l^{2}\left(R-2d\right)m\omega^{2}
\end{array}
\end{equation}
and the leading order in $\frac{d}{R}$ is
\begin{equation}
C_{1}^{pert}=-\frac{R^{2}}{4d}l^{2}m\omega^{2}.\label{eq:C1_ex}
\end{equation}
We calculate now $C_{0}^{pert}$. Eq (\ref{eq:49c}) contains an integral
of the form
\begin{equation}
l^{2}g\int_{R-2d}^{R}\rho_{0}\left(x\right)\frac{d^{2}\rho_{0}\left(x\right)}{dx^{2}}dx\ll l^{2}g\rho_{0}\left(R-2d\right)\int_{R-2d}^{R}\frac{d^{2}\rho_{0}\left(x\right)}{dx^{2}}dx.
\end{equation}
The integral over the second derivative of $\rho_{0}$ was already
calculated in (\ref{eq:second_integral}) and we can estimate
\begin{equation}
l^{2}g\int_{R-2d}^{R}\rho_{0}\left(x\right)\frac{d^{2}\rho_{0}\left(x\right)}{dx^{2}}dx\ll\left(R-2d\right)m\omega^{2}l^{2}\rho_{0}\left(R-2d\right).
\end{equation}
Using (\ref{eq:69}) and (\ref{eq:70}), it turns out that the second
term in (\ref{eq:49c}) is negligible, hence,
\begin{equation}
C_{0}^{pert}\approx\frac{1}{4}ml^{2}\omega^{2}\label{eq:C0_ex}
\end{equation}

Now, it is possible to calculate $\Delta\rho$. Let us write the ground
state of the standard GPE as a minimum with respect to $\delta\rho$
of 
\begin{equation}
\begin{array}{ccl}
E^{GP}\left(\delta\rho\right) & \thickapprox & E^{GP}\left(0\right)+\Delta E_{p}^{GP}+\Delta E_{nl}^{GP}\end{array}.
\end{equation}
Using the previous results (\ref{eq:cor_Ep}),(\ref{eq:dnlGP12}),(\ref{eq:49b})
and (\ref{eq:c1GP}) we end up with
\begin{equation}
E^{GP}\left(\delta\rho\right)=E^{GP}\left(0\right)+\frac{gR^{2}}{2d}\delta\rho^{2}.\label{eq:variation}
\end{equation}
If we repeat this calculation for the modified GPE (\ref{eq:2Deltas GP}),
we should add the term (\ref{eq:dEnl_pert}) to (\ref{eq:variation}).
According to (\ref{eq:C1_ex}) and (\ref{eq:C0_ex}), 
\begin{equation}
\Delta E_{nl}^{pert}=\frac{1}{4}ml^{2}\omega^{2}-\frac{R^{2}}{4d}l^{2}m\omega^{2}\delta\rho
\end{equation}
The resulting equation for the energy is
\begin{equation}
E\left(\delta\rho\right)=E^{GP}\left(0\right)+\frac{1}{4}ml^{2}\omega^{2}-\frac{R^{2}}{4d}l^{2}m\omega^{2}\delta\rho+\frac{gR^{2}}{2d}\left(\delta\rho\right)^{2}
\end{equation}
The minimum is found for
\begin{equation}
\delta\rho=\Delta\rho=-\frac{-\frac{1}{4d}l^{2}m\omega^{2}R^{2}}{2\left(\frac{gR^{2}}{2d}\right)}=\frac{1}{4g}l^{2}m\omega^{2}.\label{eq:101}
\end{equation}
This result is in agreement with (\ref{eq:droh}) found numerically.

The energy of the modified GPE for $\delta\rho=\Delta\rho$ is
\begin{equation}
\begin{array}{ccl}
E\left(\Delta\rho\right) & = & E^{GP}\left(0\right)+\frac{gR^{2}}{2d}\Delta\rho^{2}+\frac{1}{4}ml^{2}\omega^{2}-\frac{R^{2}}{4d}l^{2}m\omega^{2}\Delta\rho\\
 & = & E^{GP}\left(0\right)+\frac{1}{4}ml^{2}\omega^{2}-\frac{1}{32g}l^{4}m^{2}\omega^{4}
\end{array},
\end{equation}
which agrees (in first order in $l^{2}$) with (\ref{eq:tot_corE}).

\bibliographystyle{h-physrev}
\addcontentsline{toc}{section}{\refname}\bibliography{referances}
\selectlanguage{english}%

\end{document}